\documentclass[pdflatex,sn-mathphys-num]{sn-jnl}

\usepackage{graphicx}%
\usepackage{multirow}%
\usepackage{amsmath,amssymb,amsfonts}%
\usepackage{amsthm}%
\usepackage{mathrsfs}%
\usepackage[title]{appendix}%
\usepackage{xcolor}%
\usepackage{textcomp}%
\usepackage{manyfoot}%
\usepackage{booktabs}%
\usepackage{algorithm}%
\usepackage{algorithmicx}%
\usepackage{algpseudocode}%
\usepackage{listings}%
\usepackage{verbatim}
\usepackage{subcaption}
\usepackage{booktabs}
\usepackage{colortbl}
\usepackage{array}

\theoremstyle{thmstyleone}%
\theoremstyle{thmstyletwo}%

\theoremstyle{thmstylethree}%

\begin{document}

\title[Article Title]{QCS-ADME: Quantum Circuit Search for Drug Property Prediction with Imbalanced Data and Regression Adaptation}


\author*[1]{\fnm{Kangyu} \sur{Zheng}}\email{zhengk5@rpi.edu}

\author[1]{\fnm{Tianfan} \sur{Fu}}\email{futianfan@gmail.com}

\author[1]{\fnm{Zhiding} \sur{Liang}}\email{liangz9@rpi.edu}

\affil*[1]{\orgdiv{Department of Computer Science}, \orgname{Rensselaer Polytechnic Institute}, \orgaddress{\street{110 8th St}, \city{Troy}, \postcode{12180}, \state{New York}, \country{United State}}}


\abstract{The biomedical field is beginning to explore the use of quantum machine learning (QML) for tasks traditionally handled by classical machine learning, especially in predicting ADME (absorption, distribution, metabolism, and excretion) properties, which are essential in drug evaluation. However, ADME tasks pose unique challenges for existing quantum computing systems (QCS) frameworks, as they involve both classification with unbalanced dataset and regression problems. These dual requirements make it necessary to adapt and refine current QCS frameworks to effectively address the complexities of ADME predictions.\\
We propose a novel training-free scoring mechanism to evaluate QML circuit performance on imbalanced classification and regression tasks. Our mechanism demonstrates significant correlation between scoring metrics and test performance on imbalanced classification tasks. Additionally, we develop methods to quantify continuous similarity relationships between quantum states, enabling performance prediction for regression tasks. This represents a novel training-free approach to searching and evaluating QCS circuits specifically for regression applications. \\
Validation on representative ADME tasks—eight imbalanced classification and four regression—demonstrates moderate correlation between our scoring metrics and circuit performance, significantly outperforming baseline scoring methods that show negligible correlation.}




\maketitle
\begin{figure*}[h]
  \centering
  \includegraphics[width=\linewidth]{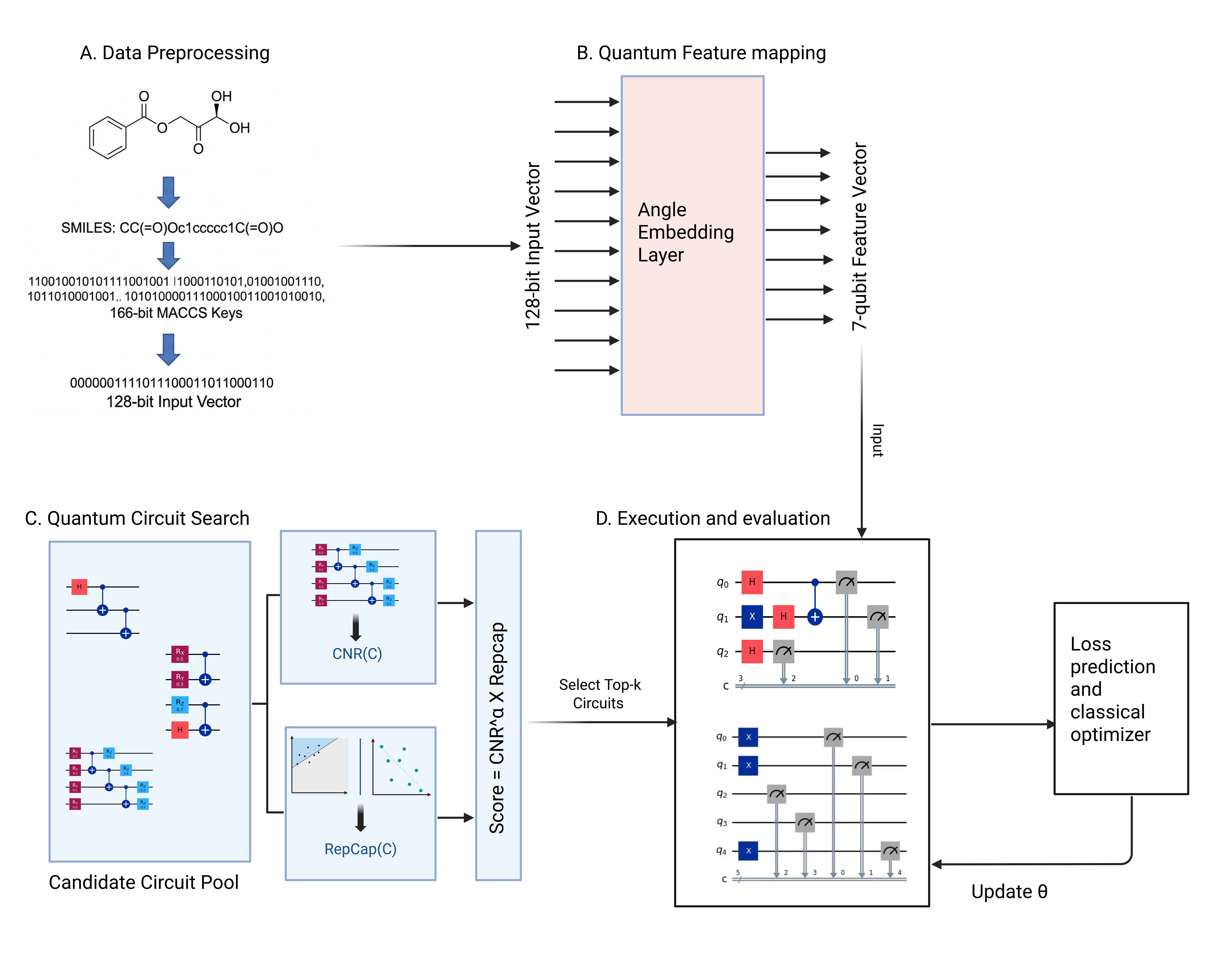}
  \caption{Overview of QCS-ADME workflow. The pipeline proceeds from A) converting SMILES string to classical bitstring, to B) mapping classical bitstring to qubit space via quantum embeeding, to C) generate candidate circuits and compute their fidelity and RepCap metrics, to D) Select the top circuits and train them.}
  \label{fig:overview}
\end{figure*}
\section{Introduction}\label{sec1}

Quantum Machine Learning (QML) has gained significant attention for its potential applications on Noisy Intermediate-Scale Quantum (NISQ) hardware \cite{Preskill_2018}, demonstrating promising results across various domains \cite{Huang_2021, Huang_2021_E, silver2023mosaiqquantumgenerativeadversarial, Huang_2022, Liu_2021}. QML architectures differ fundamentally from classical machine learning models in their use of data embedding gates, where gate selection and placement significantly impact performance, and in their gradient computation methods, which employ techniques like parameter-shift rules \cite{Schuld_2019} rather than traditional backpropagation. These unique characteristics make manual circuit design particularly challenging, often resulting in suboptimal performance. \\
\textbf{Quantum Circuit Search (QCS)} \cite{Zhang_2022, Wang_2022, Du_2022}, which is inspired from classical Neural Architecture Search (NAS) \cite{li2019randomsearchreproducibilityneural}, has emerged as a methodology for discovering high-performance, noise-robust circuits to tackle the above challenges. State-of-the-art QCS approaches~\cite{Wang_2022, 10.1145/3620665.3640354} generate circuit candidates by employing diverse algorithmic designs and evaluating performance metrics, taking into account hardware specifications and constraints. However, these methods have largely been benchmarked on simple, balanced datasets such as MNIST and Moons, limiting their applicability to complex, real-world problems. This limitation is particularly acute in the biomedical field, where predicting ADME properties is crucial for drug evaluation. ADME tasks present unique challenges that existing QCS frameworks: they frequently involve highly imbalanced classification datasets and complex regression targets.\\
While advancements like Quantum-SMOTE \cite{Mohanty2025} offer data-level solutions for imbalance, they rely on computationally expensive quantum swap tests. Within a QCS framework which requires rapid evaluation of thousands of candidates, such approaches introduce computation overhead. Consequently, there is currently no QCS scoring mechanism capable of effectively handling the dual challenges of regression topology and class imbalance found in pharmacological data. \\
To bridge this gap, we propose QCS-ADME, a novel framework that adapts QCS for biomedical property prediction. We introduce two key methodological enhancements: a weighted matrix approach for unbalanced classification tasks that amplifies minority class representation, and a modified representational capacity score for continuous similarity comparison in regression tasks. We validate these improvements on representative ADME tasks—eight unbalanced classification problems and four regression problem—demonstrating improved correlation between our enhanced scoring methods and test performance compared to traditional approaches. \\
In summary, our work not only adopts the fundamental search strategy from previous QCS methods, but also mitigates the limitations that make previous QCS methods ineffective for biomedical ADME tasks. Detailed novelty and contributions are listed as follows:  \\  
1. We develop a novel training-free scoring mechanism capable of handling regression tasks. Unlike prior methods that maximize class separability using discrete matching, our Gaussian-weighted RepCap generalizes the metric from discrete to continuous space. This encourages the quantum circuit to learn a geometry in Hilbert space that is isomorphic to the continuous topology of regression targets. \\ 
2. We introduce a density-aware weighting matrix $R_w$ to address class imbalance. This formulation modifies the search landscape to penalize circuits that collapse onto majority classes, correcting a critical failure point where standard QCS metrics yield high scores despite poor minority class performance.  \\
3. We provide the application of QCS to ADME property prediction. Furthermore, we conduct a comprehensive sim-to-real analysis, quantifying the performance gap between noiseless simulation and real hardware execution.

\section{Background}
\subsection{Quantum Circuit Search (QCS)}
Quantum Circuit Search (QCS) is a systematic approach that uses machine learning or optimization-driven techniques to automate the design of quantum circuits. Inspired by its classical counterpart, Neural Architecture Search (NAS), any QCS framework can be broken down into three core components: a \textbf{search space} that defines all possible circuits, a \textbf{search strategy} to navigate this space, and a \textbf{performance evaluation method} to assess the quality of candidate circuits. The search strategy is the engine of a QCS framework. Three dominant paradigms have emerged in this domain: Reinforcement Learning approach \cite{patel2024curriculumreinforcementlearningquantum, Zhu2023} frames circuit design as a sequential decision-making process where a software agent learns a policy for placing gates to maximize a cumulative reward signal; Evolutionary Algorithms \cite{e25010093, Zhang2023, e27070733} approach evolve circuit populations through genetic operators like crossover and mutation; Differentiable Search \cite{HE2024106508, chen2024differentiablequantumarchitecturesearch} relaxes discrete gate choices into continuous parameters amenable to gradient-based optimization. \\
The single greatest obstacle in QCS is the prohibitive cost of performance evaluation. Fully training each candidate circuit to assess its quality is computationally intractable. One plausible paradigm is training-free QCS \cite{10.1145/3620665.3640354}. This approach bypasses costly training entirely by using efficiently computable proxies to predict a circuit's performance. Instead of measuring the final outcome, these proxies analyze a circuit's intrinsic properties that are precursors to good performance, such as its expressibility, its topological complexity, or the features of its cost function landscape.
\subsection{ADME Properties}
For a small-molecule drug to be effective and safe, it must successfully navigate a complex journey through the body. This journey is characterized by a set of pharmacokinetic properties known as ADME, which stands for \textbf{Absorption, Distribution, Metabolism, and Excretion}. These properties sequentially dictate how a drug is first absorbed into the bloodstream from its site of administration, then distributed from the blood to various tissues to reach its site of action. Subsequently, the body metabolizes the drug, chemically breaking it down, before it is finally excreted along with its metabolites. \\
A drug candidate must possess a balanced and favorable ADME profile to exert its therapeutic effect without causing toxicity. In fact, a poor ADME profile is one of the most prominent reasons for the failure of drug candidates during expensive and lengthy clinical trials. Therefore, the ability to accurately predict a compound's ADME properties from its chemical structure early in the drug discovery process is crucial. Early and reliable ADME profiling allows researchers to prioritize candidates with a higher likelihood of success, saving significant time and resources. \cite{velez-arce2024signals}

\section{Method}\label{sec2}
To address the requirements of ADME Property Prediction, which encompasses both regression tasks and classification tasks with imbalanced datasets, we enhanced the QCS framework mainly in two key aspects: circuit performance evaluation on imbalanced datasets and regression task handling. The overall workflow of our proposed QCS-ADME framework is illustrated in Figure \ref{fig:overview}. The pipeline consists of four distinct phases:
\begin{itemize}
    \item \textbf{A. Data Preprocessing:} The pipeline begins by converting chemical structures into SMILES strings, which are then encoded into 166-bit MACCS keys. These are further processed to form a fixed-length 128-bit binary input vector.
    \item \textbf{B. Quantum Feature Mapping:} The classical 128-bit input vector is mapped into the quantum Hilbert space using an Angle Embedding Layer, resulting in a 7-qubit feature vector.
    \item \textbf{C. Quantum Circuit Search:} We evaluate a pool of candidate circuits using a composite scoring metric. This involves calculating the Clifford Noise Resilience (CNR) for hardware efficiency and the Representational Capacity (RepCap). The RepCap calculation incorporates our novel weighted matrix $R_w$ for imbalanced classification and Gaussian similarity for regression tasks. The final ranking score is computed as Score = $\text{CNR}^{\alpha_{\text{CNR}}} \times \text{RepCap}$.
    \item \textbf{D. Execution and Evaluation:} The top-$k$ circuits with the highest scores are selected for execution. The parameters $\theta$ are iteratively updated via a classical optimizer to minimize the prediction loss.
\end{itemize}

\noindent The structure of this section is as follow: We first outlines Elivagar's original circuit scoring methodology, then we introduce our proposed improved structure to accommodate imbalanced classification and regression tasks.
\subsection{Circuit Generation Scoring}
The circuit evaluation process consists of two main stages: circuit generation and scoring. Initially, the system generates candidate circuits based on device topology and noise-related parameters. We follow the basic framework from Elivagar \cite{10.1145/3620665.3640354}, each candidate circuit is evaluated using two predictive metrics: Clifford Noise Resilience (CNR) and Representational Capacity (RepCap). 
CNR is a training-free proxy for circuit fidelity that predicts noise robustness primarily from a circuit’s structural characteristics (e.g. depth, entangling pattern, two-qubit gate exposure) by evaluating efficiently simulable Clifford replicas of the candidate circuit, while RepCap quantifies the discriminative power of the circuit by measuring both intra-class similarities and inter-class separation in the final quantum states. These scores are combined to produce a final metric, where higher values indicate greater potential for superior performance. \\
The original RepCap scoring mechanism assumes balanced class distributions, as it requires sufficient samples to effectively distinguish between classes. To enable the capability of imbalanced classification and regression tasks common in ADME property prediction, we introduced two key steps: a weighted matrix to account for class imbalance and a revised RepCap calculation methodology for regression tasks.
\subsection{Handling Class Imbalance}
The RepCap metric evaluates circuit performance by measuring class separation in the final quantum state. A circuit receives a higher RepCap score when it maximizes inter-class separation while minimizing intra-class distances. The original RepCap formulation is defined as:
\begin{equation}
\text{RepCap}(C) = 1 - \frac{\|R_c - R_{ref}\|^2_2}{2 \cdot n_c \cdot d_c^2}
\end{equation}
Here, $d_c$ denotes the number of samples and $n_c$ represents the number of classes. $R_{ref}$ is a $d \times d$ reference matrix representing ideal circuit predictions, where $R_{ref}(i,j) = 1$ if samples $i$ and $j$ belong to the same class ($y_i = y_j$) and 0 otherwise. The matrix $R_c$ is also of dimension $d \times d$ and quantifies the similarity between quantum states of data points $i$ and $j$. Formally, we define the elements of $R_c$ as the fidelity between the variational states in Hilbert space, which constitutes the empirical Quantum Kernel $K_Q$:
\begin{equation}
    R_c(i, j) \equiv K_Q(x_i, x_j) = |\langle \psi(x_i) | \psi(x_j) \rangle|^2
\end{equation}
By minimizing the Frobenius distance between this kernel matrix and the ideal reference $R_{ref}$, RepCap maximize the alignment between kernel and the target. This ensures the selected circuit naturally maps data into a geometry that reflects the class structure, reducing the burden on the classical optimizer and improving generalization potential. \\
While this formulation performs effectively on balanced datasets, it exhibits limitations when handling imbalanced class distributions. In balanced scenarios, the reference matrix contains equal representation for each class. However, when certain classes have significantly more samples than others, the RepCap metric may assign high scores to circuits that excel at distinguishing majority classes while performing poorly on minority classes.
To address this limitation, we incorporate classical imbalanced classification techniques by introducing a weight matrix to the reference matrix, thereby amplifying the importance of minority classes. This modification ensures that circuits must develop robust representations for all classes, regardless of their distribution in the training data, to achieve high scores. The enhanced RepCap formulation is:
\begin{equation}
\text{RepCap}(C) = 1 - \frac{\|R_c - R_w \otimes R_{ref}\|^2_2}{2 \cdot n_c \cdot d_c^2},
\end{equation}
We define the weight matrix $R_w$ as $R_w(i,j) = w_i \cdot w_j$, where $w_i$ is the inverse class frequency associated with sample $i$. Specifically, $w_i = \frac{N}{N_c}$, where $N$ is total samples and $N_c$ is the count of samples in class $c$. This ensures that pairs of minority-class samples contribute significantly more to the RepCap score, forcing the circuit to distinguish them.
\subsection{Regression Task Adaptation}
For regression tasks, we maintain the fundamental structure of the classification approach while introducing key modifications to handle continuous outputs. The first significant modification involves constructing a continuous reference matrix using a Gaussian similarity function to quantify the relationship between target values:
\begin{equation}
R_{ref} = \exp\left(-\frac{\|y_i - y_j\|_2^2}{2\sigma^2}\right),
\end{equation}
where $\|y_i - y_j\|_2^2$ represents the squared Euclidean distance between target values, and $\sigma$ serves as a bandwidth parameter controlling the decay rate of similarity with respect to distance. This formulation effectively treats $R_{ref}$ as a Radial Basis Function (RBF) kernel on the labels, encoding a smoothness prior common in kernel regression: pairs with similar targets should remain close in the representation space. Consequently, aligning the quantum similarity matrix $R_c$ with the RBF label kernel $R_{ref}$ promotes a specific geometry where increasing Euclidean distance in the label space corresponds to increasing Fubini distance in the Hilbert space.\\
The second modification addresses the quantum state similarity computation. While retaining the basic quantum state similarity calculation framework, we incorporate continuous weighting based on target values. This adaptation is essential because conventional quantum state similarity measurements only capture state relationships in Hilbert space, without considering the proximity of corresponding target values. By introducing target value-dependent Gaussian weighting factors, we ensure the quantum circuit learns to map inputs with similar target values to correspondingly similar quantum states.\\
These adaptations enable our modified RepCap metric to effectively evaluate circuit performance on regression tasks, providing a more accurate assessment of each candidate circuit's capabilities. The resulting score more precisely reflects the circuit's ability to maintain the continuous relationships inherent in regression problems while preserving the quantum state fidelity considerations of the original framework.

\section{Experiments}\label{sec3}

\subsection{Dataset Selection and Preprocessing}
From the TDC \cite{velez-arce2024signals} ADME and Tox groups, we selected eight classification tasks: Human Intestinal Absorption (\verb|HIA_Hou|) \cite{hou2007adme}, P-glycoprotein Inhibition (\verb|Pgp_Broccatelli|) \cite{doi:10.1021/jm101421d}, Bioavailability (\verb|Bioavailability_Ma|) \cite{MA2008677}, Blood-Brain Barrier (\verb|BBB_Martins|) \cite{martins2012bayesian}, CYP2C9/CYP2D6 Substrate (\verb|CYP2C9_Substrate_CarbonMangels|, \verb|CYP2D6_Substrate_CarbonMangels|) \cite{https://doi.org/10.1002/minf.201100069}, Human ether-à-go-go related gene blockers (\verb|hERG|) \cite{doi:10.1021/acs.molpharmaceut.6b00471}, Drug Induced Liver Injury (\verb|DILI|) \cite{doi:10.1021/acs.jcim.5b00238}. And four regression tasks: Volumn of Distribution at steady state (\verb|VDss_Lombardo|) \cite{doi:10.1021/acs.jcim.6b00044}, Half Life of a drug (\verb|Half_Life_Obach|) \cite{obach2008trend}, Drug clearance (\verb|Clearance_Hepatocyte_AZ|, \verb|Clearance_Microsome_AZ|) \cite{DI2012441}. Note that the classification tasks here have an imbalance positive to negative ratio from 3:1 to 6:1.  \\
To prepare these datasets for our QML circuits, we implemented several preprocessing steps. First, we transformed the SMILES strings into Molecular ACCess Systems keys fingerprints (MACCS) \cite{doi:10.1021/ci010132r}. MACCS fingerprints are 166-bit strings where each bit indicates the presence (1) or absence (0) of specific molecular structural features. We utilized the first 128 bits to accommodate our quantum system's requirements. For the classification task, we remapped all 0 labels to -1. For the regression task, we normalized the target values in both training and testing sets to the range [-1, 1].

\subsection{Quantum Circuit Methodology}
\label{sec:quantum-methods}
\begin{figure}[t]
    \centering
    \includegraphics[width=1.0\linewidth]{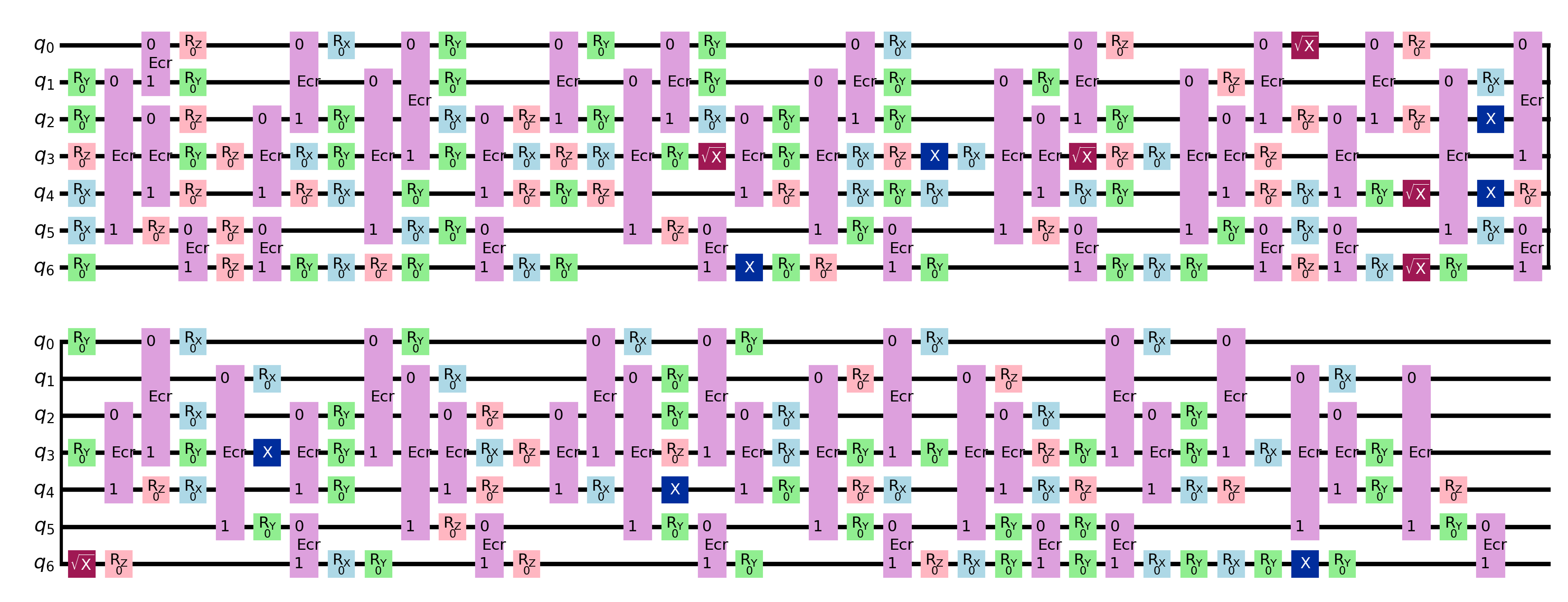} 
    \caption{One candidate quantum circuit found for the \texttt{BBB\_Martins} task. The circuit is transpiled for the IBM Rensselaer processor, utilizing 7 qubits with angle encoding and variational layers containing parameterized rotations and hardware-native ECR entangling gates.}
    \label{fig:circuit_diagram}
\end{figure}
\textbf{Circuit Architecture and Encoding}: We employ device-aware variational quantum circuits optimized for the 127-qubit IBM Rensselaer superconducting processor. For the ADME tasks, we utilize a subsystem of $n=7$ qubits. Given a molecular descriptor vector $x \in \mathbb{R}^{128}$, we employ an Angle Encoding strategy where each feature $x_j$ parameterizes a single-qubit rotation. Specifically, the data embedding layer consists of a sequence of $R_Y$ rotations:
\begin{equation}
    R_Y(x_j)|0\rangle = \cos\left(\frac{x_j}{2}\right)|0\rangle + \sin\left(\frac{x_j}{2}\right)|1\rangle
\end{equation}
The full circuit integrates 128 such data-encoding gates interleaved with variational parameters, resulting in a dense encoding of the feature space. The ansatz contains 32 trainable parameters optimized during the learning process. Figure \ref{fig:circuit_diagram} illustrates a representative optimized circuit architecture found by our search algorithm. This diagram explicitly visualizes the native gate decomposition used during execution on the hardware.

\noindent \textbf{Noise-Aware Circuit Construction}: Candidate circuits for the search space are generated via a biased random sampling strategy that incorporates real-time device calibration data. The sampling probability for placing a gate on specific qubits is weighted by:
\begin{itemize}
    \item Qubit coherence times ($T_1$, $T_2$)
    \item Two-qubit gate fidelities (CNOT/ECR error rates)
    \item Measurement readout fidelities
    \item Physical hardware connectivity constraints
\end{itemize}
This probability distribution biases the construction toward higher-fidelity qubits and links, effectively pruning the search space of hardware-inefficient architectures before training begins.

\noindent \textbf{Measurement and Prediction}: The final prediction is derived from the expectation value of the Pauli-$Z$ operator, measured on a single readout qubit $q_{out}$ selected for optimal readout fidelity:
\begin{equation}
    f(x,\theta) = \langle \psi(x,\theta) | Z_{q_{out}} | \psi(x,\theta) \rangle \in [-1,1]
\end{equation}
For binary classification tasks, the discrete label is assigned via $\operatorname{sign}(f(x,\theta))$.

\subsection{Training and Evaluation Configuration}
Our training configuration largely follows the Elivagar framework, with necessary adaptations for our specific tasks. We utilized the IBM Rensselaer quantum device architecture and its corresponding noise model for candidate circuit generation. To accommodate our 128-length input features, we configured the system with 7 qubits and employed 32 Clifford replicas for CNR computation. We generated 250 candidate circuits, which were utilized across both classification and regression tasks. \\
For both tasks, we trained the circuits for 200 epochs with a batch size of 256, using the Adam optimizer with a learning rate of 0.01. The training objective was to minimize the loss between circuit predictions and the corresponding labels or target values in the training set. All training procedures were executed on noiseless simulators using the TorchQuantum \cite{Wang_2022} framework, with computations performed on a 16-core AMD Ryzen Threadripper PRO 5955WX processor. \\
In evaluation, we run all the circuits once for both tasks and calculate the final score with the same equation as used by Elivagar.
\begin{equation}
    \text{Score}(C) = \text{CNR}(C)^{\alpha_{\text{CNR}}} \times \text{RepCap}(C),
\end{equation}
where we choose $\alpha_{\text{CNR}} = 0.25$. \\
Beside evaluating on classical computer, we also evaluate four classification tasks and two regression tasks' circuits on real IBM Rensselaer quantum hardware. Both of these experiments results are shown in Section \ref{sec4} and Appendix \ref{secA1}.

\section{Results}\label{sec4}
We evaluated performance using metrics tailored to each task. For both classification and regression, we calculated the Spearman correlation between the final score and the test Mean Squared Error. For classification tasks, we used Accuracy, F1-score, and the Area Under the Precision-Recall Curve (PR-AUC). For regression tasks, we additionally reported the MSE loss itself. All our implementation code can be found at \url{https://github.com/zkysfls/quantum_admet}.\\

\subsection{Classification Tasks Performance}

\begin{figure*}[htbp]
\centering
  \begin{subfigure}{0.47\linewidth} 
  \centering
    \includegraphics[width=\linewidth]{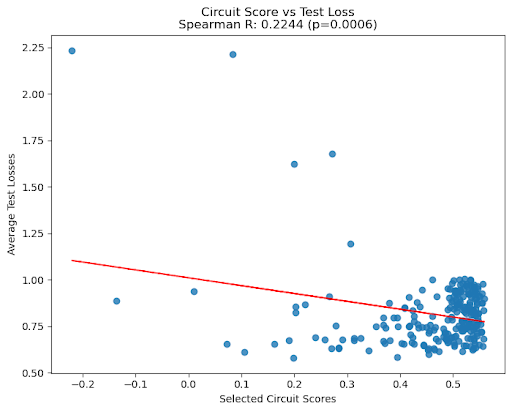} 
    \caption{\small QCS\_ADME on BBB\_Martins}
    \label{1-a}
  \end{subfigure}
  \begin{subfigure}{0.47\linewidth}
  \centering
    \includegraphics[width=\linewidth]{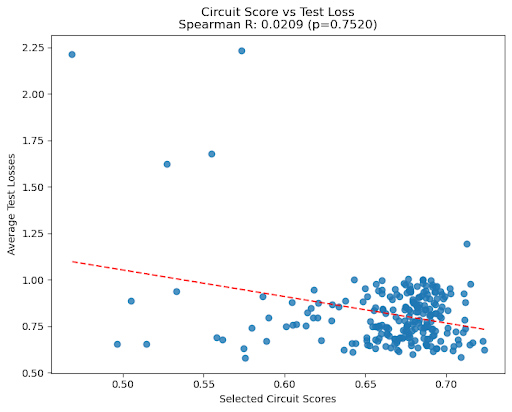} 
    \caption{\small Elivagar on BBB\_Martins}
    \label{1-b}
  \end{subfigure}
  \vspace{1.5em}
  \begin{subfigure}{0.47\linewidth} 
  \centering
    \includegraphics[width=\linewidth]{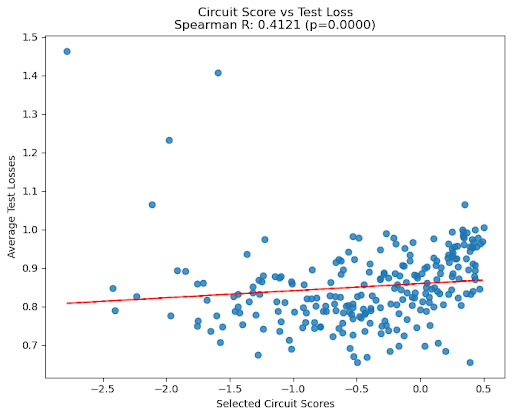} 
    \caption{\small QCS\_ADME on Hia\_Hou}
    \label{1-c}
  \end{subfigure}
  \begin{subfigure}{0.47\linewidth}
  \centering
    \includegraphics[width=\linewidth]{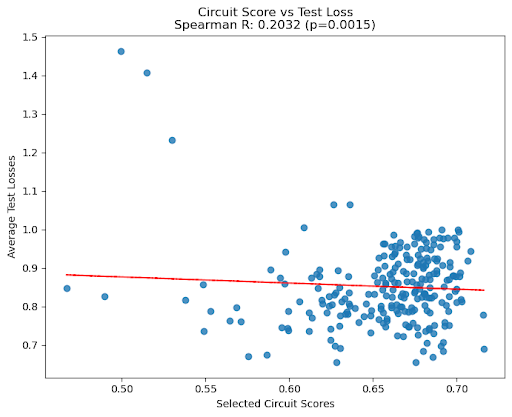} 
    \caption{\small Elivagar on Hia\_Hou}
    \label{1-d}
  \end{subfigure}
  \caption{Spearman correlation between scores and test loss on unbalanced classification tasks BBB\_Martins and Hia\_Hou. Our revised score shows a stronger correlation compared to the original score. } 
  \label{1}
\end{figure*}

\begin{figure*}[htbp]
  \centering
  \includegraphics[width=\linewidth]{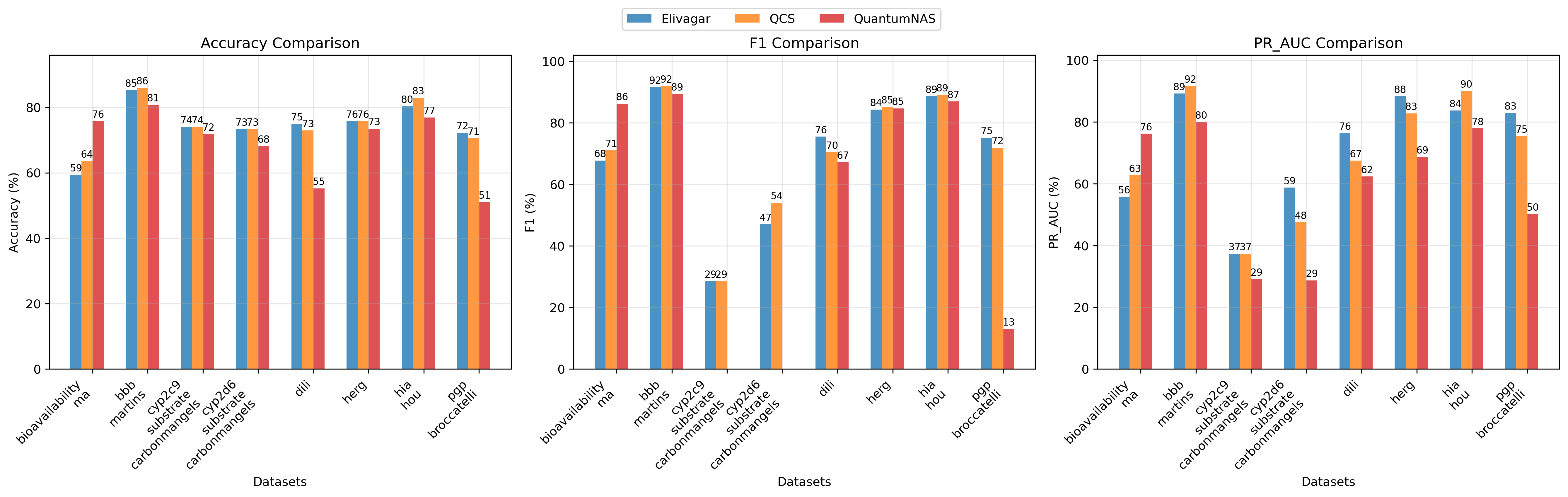}
  \caption{The overview comparison between our proposed QCS method with other Quantum Circuit Search methods. }
  \label{2}
\end{figure*}

Figure \ref{1} compares the performance of the original Elivagar scoring system with our enhanced scoring method on two imbalanced classification tasks: BBB\_Martins and Hia\_Hou. As shown in Figure \ref{1-b} and Figure \ref{1-d}, the original Elivagar \cite{10.1145/3620665.3640354} final score demonstrates a weak predictive relationship with the average test loss. On the BBB\_Martins dataset, the correlation is negligible and not statistically significant ($R=0.0269, p=0.7520$), while on the Hia\_Hou dataset, it is weak but significant ($R=0.2032, p=0.0015$). In contrast, our revised scoring method shows a consistently stronger correlation. On BBB\_Martins (Figure \ref{1-a}), it achieves a Spearman correlation of ($R=0.2244, p=0.0006$), and on Hia\_Hou (Figure \ref{1-c}), it reaches a moderate correlation of ($R=0.4121, p<0.0001$). These results suggest that our modifications successfully capture circuit performance characteristics in these two imbalanced classification tasks, making our score a more reliable metric.
While Figure \ref{1} illustrates two representative datasets where QCS-ADME strengthens the score–test-loss correlation, the effect is dataset-dependent. We therefore report correlations across all classification datasets in Appendix Table \ref{tab:appendix_corr_cls} and analyze failure cases where correlation improvements are weak or negative in section \ref{sec:ablation}.\\
We also evaluate the performance of our quantum-generated circuits against other Quantum Circuit Search methods on all selected classification tasks. For this comparison, we choose the QML circuits ranked in the Top-25 by our revised score.
As shown in Figure \ref{2}, we compare four circuit selection methods: our proposed QCS (orange), Elivagar \cite{10.1145/3620665.3640354} (blue), QuantumNAS \cite{Wang_2022} (red). Across all three classification metrics, a clear trend emerges, with our proposed QCS method consistently demonstrating robust and high-quality performance. For a majority of the tasks—including \verb|CYP2C9_Substrate_CarbonMangels|, \verb|CYP2D6_Substrate_CarbonMangels|, \verb|BBB_Martins|, \verb|hERG|, and \verb|HIA_Hou|—the circuits selected by QCS achieve results that are either the best or highly competitive with the other leading methods. For instance, on the \verb|BBB_Martins| task, QCS leads across all three metrics, achieving an accuracy of approximately 86\%, an F1-score of approximately 92\%, and a PR-AUC of approximately 92\%. \\
Furthermore, we evaluated the performance of our quantum-generated circuits against classical machine learning models (XGBoost, Random Forest, SVM, and Logistic Regression). As shown in Table \ref{table_bbb} and Table \ref{table_hia}, the performance of the QML circuits presents a mixed picture when compared to the classical baselines. For the \verb|BBB_Martins| task, the best-performing QML circuit achieves a superior F1-score of 0.9198, outperforming all listed classical models. However, its PR-AUC of 0.9301 lags behind the classical approaches, such as Random Forest (0.9650) and XGBoost (0.9639). On the \verb|Hia_Hou| task, the performance gap is more pronounced. The best QML circuit's F1-score (0.8913) is surpassed by several classical models like XGBoost (0.9554), and its PR-AUC (0.9384) is substantially lower than the near-perfect scores achieved by all the classical baselines (all $>0.995$). These results, with detailed comparisons for additional tasks available in the Appendix, highlight that while QML circuits can be competitive on certain metrics, a significant gap often remains, particularly in PR-AUC.

\begin{table}[htbp]
\centering
  \caption{Performance comparison for BBB\_Martins.}
  \label{table_bbb}
  \begin{minipage}{\textwidth}
    \begin{tabular}{c|c|c|c}
      \toprule[0.8pt]
      \textbf{Metric} & \textbf{QML average} & \textbf{QML best} & \textbf{XGBoost} \\
      \hline
      Accuracy & \(0.7831 \pm 0.0643\) & \(0.8596\) & \(0.8794 \pm 0.0158\) \\
      F1       & \(0.8684 \pm 0.0503\) & \(0.9198\) & \(0.8798 \pm 0.0158\) \\
      PR-AUC   & \(0.8804 \pm 0.0221\) & \(0.9301\) & \(0.9639 \pm 0.0104\) \\
      \hline
    \end{tabular}
  \end{minipage}
  \par
  \vspace{1em}
  \par

  \begin{minipage}{\textwidth}
    \begin{tabular}{c|c|c|c}
      \hline
      \textbf{Metric} & \textbf{Random Forest} & \textbf{SVM} & \textbf{Logistic Regression} \\
      \hline
      Accuracy & \(0.8871 \pm 0.0157\) & \(0.8780 \pm 0.0160\) & \(0.8578 \pm 0.0179\) \\
      F1       & \(0.8795 \pm 0.0176\) & \(0.8659 \pm 0.0187\) & \(0.8599 \pm 0.0176\) \\
      PR-AUC   & \(0.9650 \pm 0.0114\) & \(0.9578 \pm 0.0118\) & \(0.9464 \pm 0.0154\) \\
\bottomrule[0.8pt]
    \end{tabular}
  \end{minipage}
\end{table}

\begin{table}[htbp]
\centering
  \caption{Performance comparison for Hia\_Hou.}
  \label{table_hia}
  \begin{minipage}{\textwidth}
    \begin{tabular}{c|c|c|c}
      \toprule[0.8pt]
      \textbf{Metric} & \textbf{QML average} & \textbf{QML best} & \textbf{XGBoost} \\
      \hline
      Accuracy & \(0.6971 \pm 0.0830\) & \(0.8291\) & \(0.9573 \pm 0.0182\) \\
      F1       & \(0.7980 \pm 0.0733\) & \(0.8913\) & \(0.9554 \pm 0.0196\) \\
      PR-AUC   & \(0.8529 \pm 0.0552\) & \(0.9384\) & \(0.9976 \pm 0.0018\) \\
      \hline
    \end{tabular}
  \end{minipage}
  \par
  \vspace{1em}
  \par

  \begin{minipage}{\textwidth}
    \begin{tabular}{c|c|c|c}
      \hline
      \textbf{Metric} & \textbf{Random Forest} & \textbf{SVM} & \textbf{Logistic Regression} \\
      \hline
      Accuracy & \(0.9056 \pm 0.0261\) & \(0.8816 \pm 0.0298\) & \(0.9408 \pm 0.0218\) \\
      F1       & \(0.8954 \pm 0.0310\) & \(0.8644 \pm 0.0367\) & \(0.9393 \pm 0.0228\) \\
      PR-AUC   & \(0.9954 \pm 0.0031\) & \(0.9964 \pm 0.0022\) & \(0.9960 \pm 0.0027\) \\
\bottomrule[0.8pt]
    \end{tabular}
  \end{minipage}
\end{table}
\subsection{Regression Task Performance}

\begin{figure*}[htbp]
\centering
  \begin{subfigure}{0.47\linewidth} 
  \centering
    \includegraphics[width=\linewidth]{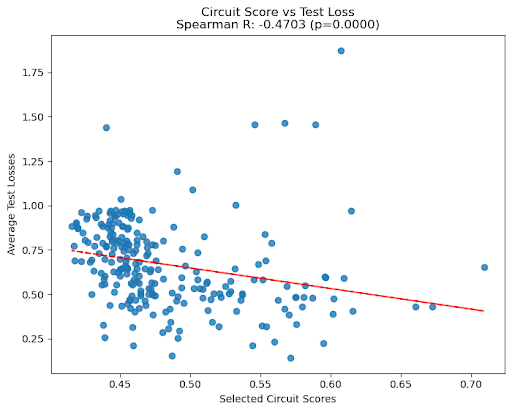} 
    \caption{\small Half\_Life\_Obach}
    \label{3-a}
  \end{subfigure}
  \begin{subfigure}{0.47\linewidth}
  \centering
    \includegraphics[width=\linewidth]{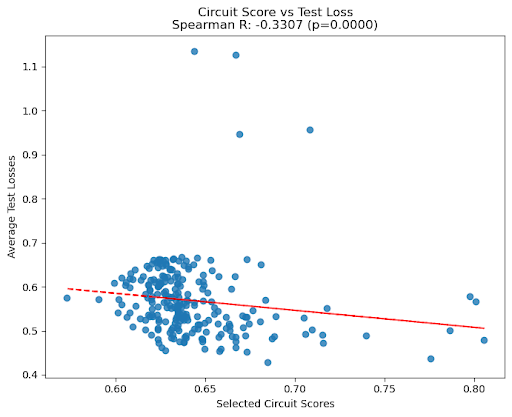} 
    \caption{\small Clearance\_Hepatocyte\_AZ}
    \label{3-b}
  \end{subfigure}
  \vspace{1.5em}
  \begin{subfigure}{0.47\linewidth} 
  \centering
    \includegraphics[width=\linewidth]{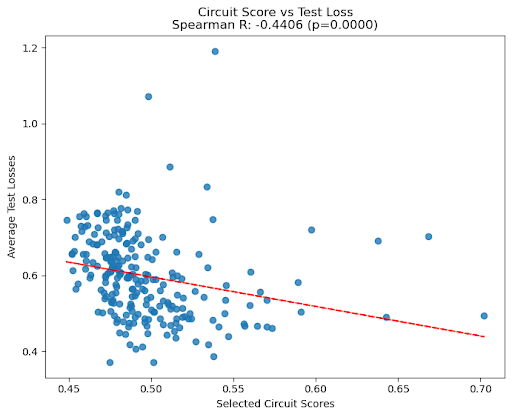} 
    \caption{\small Clearance\_Microsome\_AZ}
    \label{3-c}
  \end{subfigure}
  \begin{subfigure}{0.47\linewidth}
  \centering
    \includegraphics[width=\linewidth]{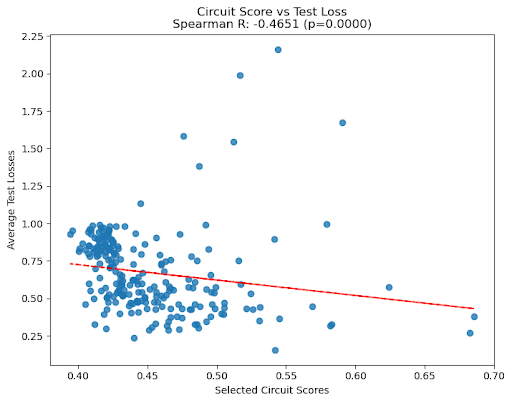} 
    \caption{\small VDss\_Lombardo}
    \label{3-d}
  \end{subfigure}
  \caption{\small Spearman correlation between scores and test loss on regression tasks.} 
  \label{3}
\end{figure*}

Figure \ref{3} showcases the relationship between our proposed circuit scores and final model performance, measured by average test loss, across four distinct regression tasks. The scatter plots for the \verb|VDss_Lombardo|, \verb|Half_Life_Obach|, \verb|Clearance_Hepatocyte_AZ|, \verb|Clearance_Microsome_AZ| tasks consistently demonstrate a statistically significant negative correlation between the two variables. This is the desired outcome, as it confirms that circuits with higher scores are strongly associated with lower test losses, indicating better predictive performance. The Spearman correlation coefficients (R) are moderate and negative, ranging from -0.3367 to -0.4881, and are all highly significant ($p<0.0001$). These results validate our scoring metric as an effective tool for identifying superior quantum circuits for regression problems.\\

\begin{figure*}[htbp]
  \centering
  \includegraphics[width=\linewidth]{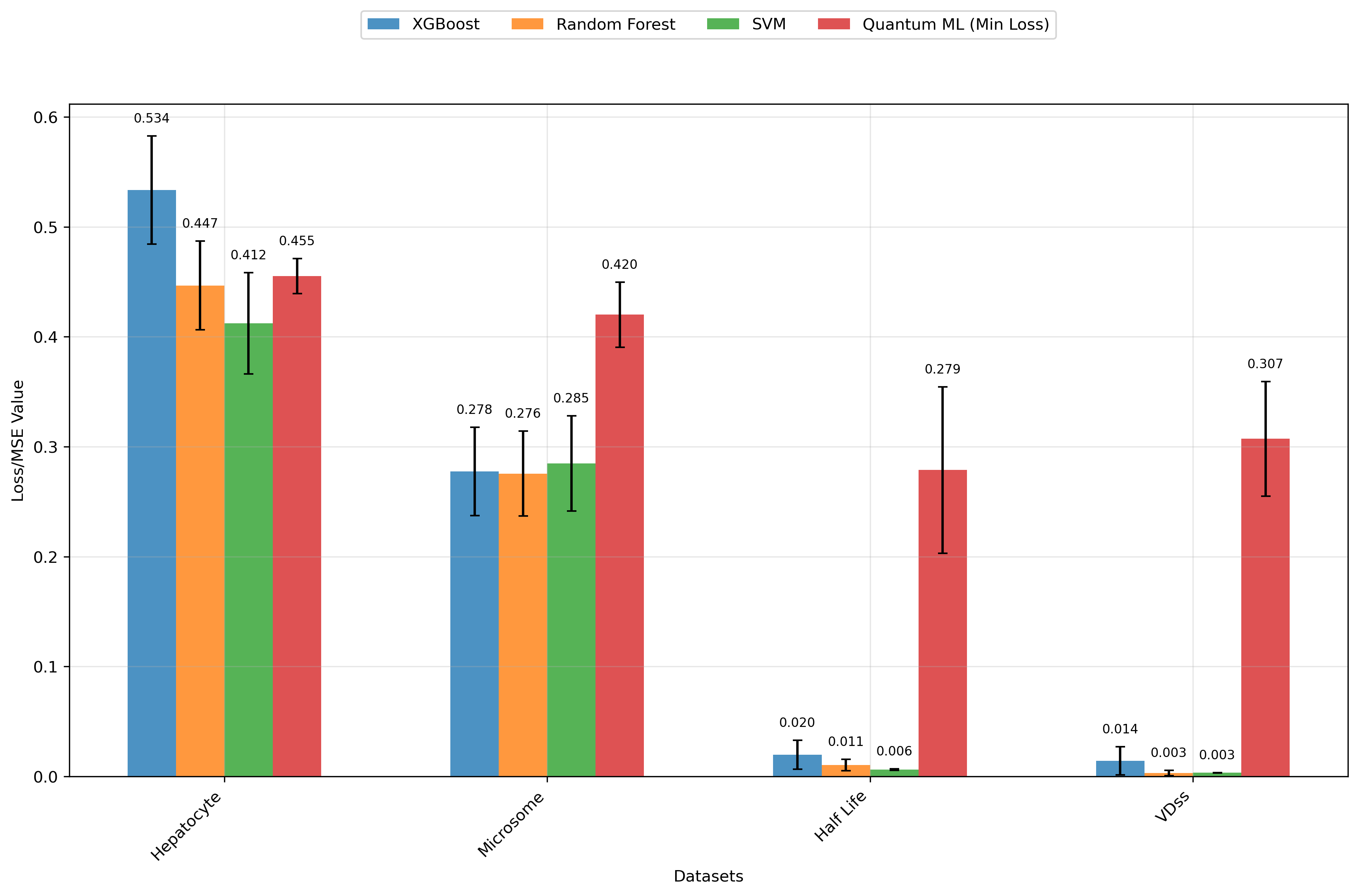}
  \caption{The comparison between our proposed QCS method with classical ML models on regression tasks}
  \label{4}
\end{figure*}
Finally, we evaluate the regression performance of our generated QML circuits against classical machine learning models. We follow the classification comparsion setting that take the Top-25 circuits based on our score and compare to the same set of classical ML models (XGBoost, Random Forest, and SVM). The results in figure \ref{4} reveal a significant performance gap, with the classical models consistently achieving substantially lower MSE values than the QML approach in all evaluated cases. For instance, on the \verb|Clearance_Hepatocyte_AZ| and \verb|Clearance_Microsome_AZ| tasks, the QML model's MSE of 0.455 and 0.420, respectively, is markedly higher than its classical counterparts. This disparity becomes even more pronounced on the \verb|Half_Life_Obach| and \verb|VDss_Lombardo| tasks, where classical models achieve exceptionally low MSEs ( 0.011 for Random Forest and 0.003 for SVM), while the QML model's loss remains significantly higher at 0.279 and 0.307, respectively. These findings underscore a key challenge: while our framework can effectively generate and evaluate quantum circuits, bridging the performance gap to established classical regression algorithms remains a critical area for future research.
\subsection{Real hardware Performance}
\begin{figure*}[htbp]
  \centering
  \includegraphics[width=\linewidth]{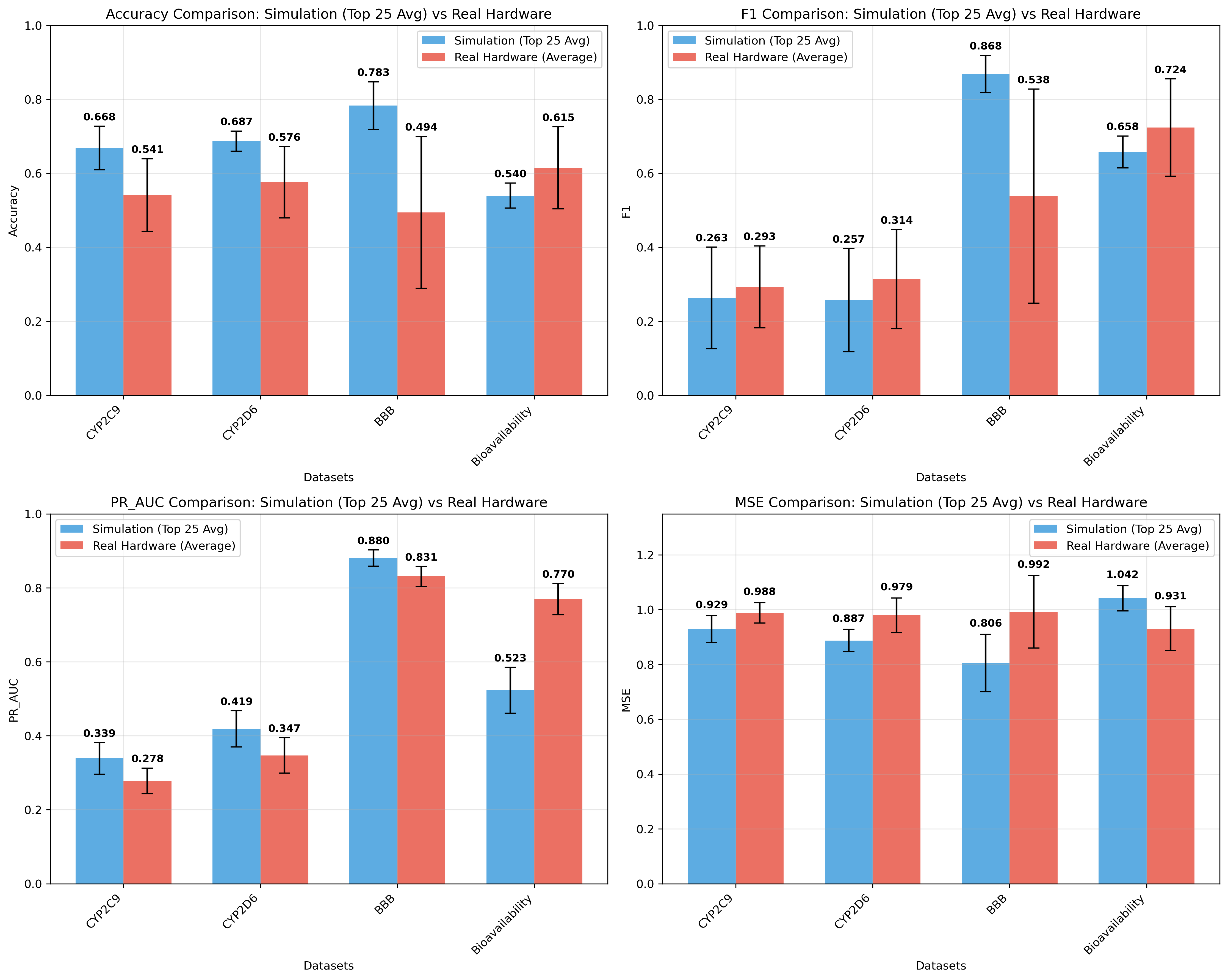}
  \caption{The comparison between simulation and real hardware performances of selected classification tasks}
  \label{5}
\end{figure*}
To further investigate our proposed method's performance, we chose four classification tasks and two regression tasks to run on IBM Rensselaer, which is the hardware configuration we chose to generate our circuits. Figure \ref{5} shows the overview comparison of simulation to real hardware on classification tasks. We could see that a general degradation in performance is observed when transitioning from simulation to noisy hardware execution. This is particularly evident in the \verb|BBB_Martins| task, where the average accuracy drops from 0.783 to 0.494 and the F1-score decreases from 0.868 to 0.538. This trend, coupled with a significant increase in performance variance on the hardware, highlights the impact of device noise. However, the \verb|Bioavailability_Ma| task presents a compelling and counter-intuitive exception. On this task, the circuits executed on real hardware consistently outperformed their simulated counterparts across all metrics. The average PR-AUC, for instance, improved from 0.523 in simulation to 0.770 on the hardware, while the Mean Squared Error advantageously decreased from 1.042 to 0.931. This suggests that for certain problem landscapes, the inherent noise of the hardware may act as a form of regularization, potentially aiding the optimization process and leading to superior generalization compared to the ideal, noiseless environment.\\
\begin{figure*}[htbp]
  \centering
  \includegraphics[width=\linewidth]{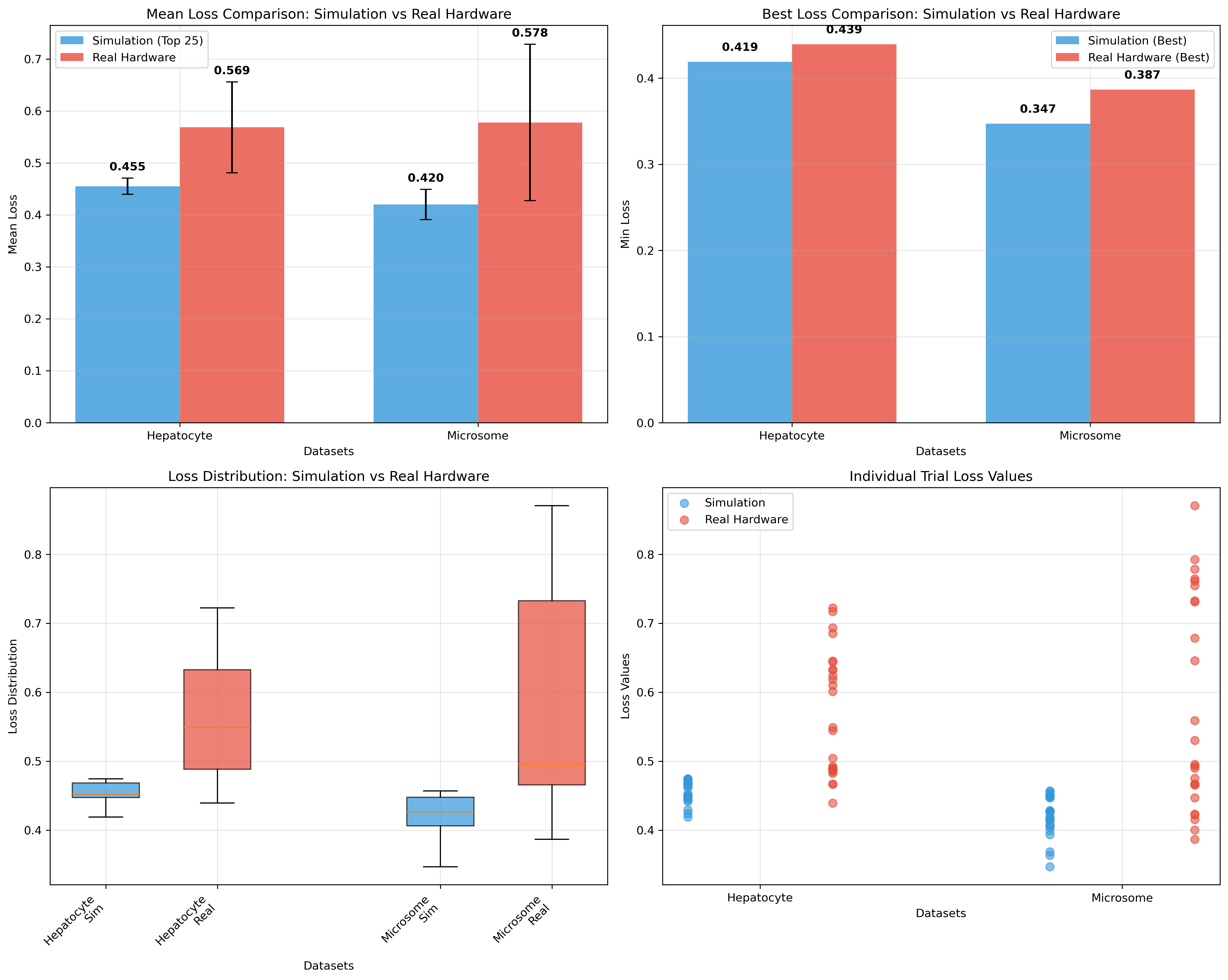}
  \caption{The comparison between simulation and real hardware performances of selected regression tasks}
  \label{6}
  \end{figure*}
For regression tasks, figure \ref{6} shows an overview comparison. The result reveals a significant performance gap between noiseless simulation and execution on real hardware. This "sim-to-real" gap is evident in both the average and best-case scenarios. The mean loss increased substantially by 24.9\% for the Hepatocyte dataset and 37.5\% for the Microsome dataset when moving from simulation to hardware. Similarly, the best-case performance was also degraded, with the minimum achievable loss increasing by 4.8\% and 11.4\% for the respective datasets. Beyond the degradation in accuracy, the performance on real hardware demonstrated markedly higher instability. This is quantified by the standard deviation of the loss, which was over five times larger on hardware for both datasets, and is visually confirmed by the wider performance distributions in the box plots. These results underscore the profound impact of hardware noise, which not only diminishes the predictive power of the models but also severely impacts the reliability and consistency of their performance on regression tasks.\\
In summary, our "sim-to-real" analysis demonstrates that hardware noise has a complex and task-dependent impact on performance. While noise was uniformly detrimental to our regression tasks, consistently increasing error and instability, its effect on classification was more nuanced. We found that for some classification tasks, noise paradoxically improved results, suggesting it can act as a potential regularizer in addition to being a performance bottleneck. This dual nature of hardware noise necessitates task-aware strategies for developing effective near-term quantum algorithms.
\subsection{Ablation Study and Failure Analysis}\label{sec:ablation}
To justify the design of the weighted matrix $R_w$, we conducted an ablation study comparing our method against the baseline Elivagar scoring which lacks imbalance adaptation. As shown in Table \ref{tab:appendix_corr_cls}, the impact of the weighting mechanism is task-dependent.\\
\textbf{Success Cases:} On the \texttt{hERG} and \texttt{Bioavailability\_Ma} datasets, the introduction of $R_w$ transformed the scoring metric from random or weakly predictive to strongly predictive. For instance, on \texttt{hERG}, the Spearman correlation improved from $R=0.058$ to $R=-0.276$, and the best-found circuit achieved an F1-score of 0.852, surpassing the baseline's 0.843. This demonstrates that for datasets where the minority class features are distinct but underrepresented, the weighted RepCap successfully guides the search toward robust architectures.\\
\\textbf{Limitations and Failure Modes:} Conversely, we acknowledge specific limitations regarding data quality and study scope. First, we observed inconsistencies in datasets such as \texttt{HIA\_Hou} and \texttt{BBB\_Martins}, where the metric exhibited a positive correlation with test loss (e.g., $R=0.412$ for \texttt{HIA\_Hou}). We hypothesize this is due to the noise profile of these specific ADME datasets. In scenarios where the minority class contains significant label noise or outliers, the weighted metric effectively rewards overfitting to the training geometry, which in result prioritize circuits that separate these outliers at the expense of learning the true decision boundary. This suggests that while $R_w$ is a powerful tool for imbalance, it requires datasets with reliable minority labels to function optimally. Second, we restricted our evaluation to TDC benchmarks rather than larger industrial datasets due to the computational constraints of simulating Quantum Circuit Search. The processing of industrial-scale feature spaces currently exceeds NISQ-era simulation capacities, necessitating future hardware advancements to bridge this gap.
\section{Conclusion}
This paper presents a QCS framework adapted for unbalanced classification and regression tasks in QML circuits. Building upon an established balanced classification QCS framework, we modified the circuit performance scoring methodology through two key innovations: a weight adjustment matrix for unbalanced tasks and a Gaussian similarity metric for continuous regression relationships. \\
Evaluation on medical task datasets demonstrates moderate correlation between our revised scoring method and final test loss. This validates our score as a more reliable predictor of model performance than previous methods, particularly for regression tasks where it showed consistent negative correlation. Circuits selected by our framework were highly competitive with those from other state-of-the-art quantum circuit search methods. However, when compared to classical models, a performance gap remains; while our QML circuits were competitive in some classification metrics, they were significantly outperformed in regression tasks. Finally, our "sim-to-real" analysis revealed that the impact of hardware noise is highly task-dependent: it was uniformly detrimental to regression performance but had a more nuanced effect on classification, even paradoxically improving results in one instance. These findings highlight both the utility of our adapted framework and the ongoing challenges of bridging the performance gap to classical algorithms and navigating the complex effects of noise in the NISQ era. \\
There are several possible directions to further improve our work. Firstly, we will focus on developing advanced methods to generate more competitive candidate circuits. Second, while our use of Clifford Noise Resilience (CNR) effectively guides the search toward robust architectures, the sim-to-real gap in regression tasks suggests the need for further mitigation. Future development could incorporate task-aware circuit pruning, where gates with negligible gradients regarding the specific regression loss are iteratively removed before hardware execution. Additionally, we plan to incorporate label-noise robustness terms into the $R_w$ formulation.

\section{Declarations}
\textbf{Funding}: This work is funded through the IBM-RPI Future of Computing Research Collaboration. \\
\textbf{Clinical trial number}: not applicable.
\clearpage
\bibliography{sn-bibliography}
\begin{appendices}

\section{Additional results}\label{secA1}
\begin{table}[htbp]
\centering
  \caption{Performance comparison for CYP2C9\_Substrate\_CarbonMangels task.}
  \label{table_cyp2c9}
  \begin{minipage}{\textwidth}
    \begin{tabular}{c|c|c|c}
      \toprule[0.8pt]
      \textbf{Metric} & \textbf{QML average} & \textbf{QML best} & \textbf{XGBoost} \\
      \hline
      Accuracy & \(0.6684 \pm 0.0590\) & \(0.7407\) & \(0.7041 \pm 0.0391\) \\
      F1       & \(0.2630 \pm 0.1373\) & \(0.4691\) & \(0.6590 \pm 0.0472\) \\
      PR-AUC   & \(0.3390 \pm 0.0426\) & \(0.4297\) & \(0.3922 \pm 0.0739\) \\
      \hline
    \end{tabular}
  \end{minipage}
  \par
  \vspace{1em}
  \par

  \begin{minipage}{\textwidth}
    \begin{tabular}{c|c|c|c}
      \hline
      \textbf{Metric} & \textbf{Random Forest} & \textbf{SVM} & \textbf{Logistic Regression} \\
      \hline
      Accuracy & \(0.6971 \pm 0.0392\) & \(0.7193 \pm 0.0374\) & \(0.7110 \pm 0.0385\) \\
      F1       & \(0.6033 \pm 0.0501\) & \(0.6024 \pm 0.0494\) & \(0.6633 \pm 0.0473\) \\
      PR-AUC   & \(0.3760 \pm 0.0670\) & \(0.3921 \pm 0.0682\) & \(0.3854 \pm 0.0723\) \\
\bottomrule[0.8pt]
    \end{tabular}
  \end{minipage}
\end{table}

\begin{table}[htbp]
\centering
  \caption{Performance comparison for CYP2D6\_Substrate\_CarbonMangels task.}
  \label{table_cyp2d6}
  \begin{minipage}{\textwidth}
    \begin{tabular}{c|c|c|c}
      \toprule[0.8pt]
      \textbf{Metric} & \textbf{QML average} & \textbf{QML best} & \textbf{XGBoost} \\
      \hline
      Accuracy & \(0.6871 \pm 0.0272\) & \(0.7333\) & \(0.7276 \pm 0.0379\) \\
      F1       & \(0.2574 \pm 0.1396\) & \(0.5405\) & \(0.7229 \pm 0.0391\) \\
      PR-AUC   & \(0.4189 \pm 0.0489\) & \(0.5174\) & \(0.5757 \pm 0.0768\) \\
      \hline
    \end{tabular}
  \end{minipage}
  \par
  \vspace{1em}
  \par

  \begin{minipage}{\textwidth}
    \begin{tabular}{c|c|c|c}
      \hline
      \textbf{Metric} & \textbf{Random Forest} & \textbf{SVM} & \textbf{Logistic Regression} \\
      \hline
      Accuracy & \(0.7407 \pm 0.0383\) & \(0.7405 \pm 0.0384\) & \(0.7190 \pm 0.0396\) \\
      F1       & \(0.7227 \pm 0.0427\) & \(0.7191 \pm 0.0437\) & \(0.7151 \pm 0.0410\) \\
      PR-AUC   & \(0.6292 \pm 0.0744\) & \(0.6390 \pm 0.0730\) & \(0.5690 \pm 0.0803\) \\
\bottomrule[0.8pt]
    \end{tabular}
  \end{minipage}
\end{table}

\begin{table}[htbp]
\centering
  \caption{Performance comparison for DILI task.}
  \label{table_dili}
  \begin{minipage}{\textwidth}
    \begin{tabular}{c|c|c|c}
      \toprule[0.8pt]
      \textbf{Metric} & \textbf{QML average} & \textbf{QML best} & \textbf{XGBoost} \\
      \hline
      Accuracy & \(0.6279 \pm 0.0534\) & \(0.7292\) & \(0.7992 \pm 0.0401\) \\
      F1       & \(0.6221 \pm 0.0725\) & \(0.7327\) & \(0.7966 \pm 0.0411\) \\
      PR-AUC   & \(0.6811 \pm 0.0757\) & \(0.8111\) & \(0.8783 \pm 0.0374\) \\
      \hline
    \end{tabular}
  \end{minipage}
  \par
  \vspace{1em}
  \par

  \begin{minipage}{\textwidth}
    \begin{tabular}{c|c|c|c}
      \hline
      \textbf{Metric} & \textbf{Random Forest} & \textbf{SVM} & \textbf{Logistic Regression} \\
      \hline
      Accuracy & \(0.8218 \pm 0.0380\) & \(0.8115 \pm 0.0384\) & \(0.8019 \pm 0.0399\) \\
      F1       & \(0.8195 \pm 0.0388\) & \(0.8094 \pm 0.0393\) & \(0.8009 \pm 0.0403\) \\
      PR-AUC   & \(0.8988 \pm 0.0365\) & \(0.8992 \pm 0.0332\) & \(0.8479 \pm 0.0465\) \\
\bottomrule[0.8pt]
    \end{tabular}
  \end{minipage}
\end{table}

\begin{table}[htbp]
\centering
  \caption{Performance comparison for hERG task.}
  \label{table_herg}
  \begin{minipage}{\textwidth}
    \begin{tabular}{c|c|c|c}
      \toprule[0.8pt]
      \textbf{Metric} & \textbf{QML average} & \textbf{QML best} & \textbf{XGBoost} \\
      \hline
      Accuracy & \(0.7206 \pm 0.0330\) & \(0.7576\) & \(0.8177 \pm 0.0323\) \\
      F1       & \(0.8244 \pm 0.0288\) & \(0.8546\) & \(0.8039 \pm 0.0367\) \\
      PR-AUC   & \(0.7953 \pm 0.0427\) & \(0.8584\) & \(0.9175 \pm 0.0280\) \\
      \hline
    \end{tabular}
  \end{minipage}
  \par
  \vspace{1em}
  \par

  \begin{minipage}{\textwidth}
    \begin{tabular}{c|c|c|c}
      \hline
      \textbf{Metric} & \textbf{Random Forest} & \textbf{SVM} & \textbf{Logistic Regression} \\
      \hline
      Accuracy & \(0.8330 \pm 0.0325\) & \(0.7947 \pm 0.0350\) & \(0.7794 \pm 0.0347\) \\
      F1       & \(0.8142 \pm 0.0386\) & \(0.7695 \pm 0.0421\) & \(0.7478 \pm 0.0427\) \\
      PR-AUC   & \(0.9130 \pm 0.0281\) & \(0.8968 \pm 0.0342\) & \(0.8343 \pm 0.0443\) \\
\bottomrule[0.8pt]
    \end{tabular}
  \end{minipage}
\end{table}

\begin{table}[htbp]
\centering
  \caption{Performance comparison for Pgp\_Broccatelli task.}
  \label{table_pgp}
  \begin{minipage}{\textwidth}
    \begin{tabular}{c|c|c|c}
      \toprule[0.8pt]
      \textbf{Metric} & \textbf{QML average} & \textbf{QML best} & \textbf{XGBoost} \\
      \hline
      Accuracy & \(0.6150 \pm 0.0502\) & \(0.7061\) & \(0.8207 \pm 0.0249\) \\
      F1       & \(0.6396 \pm 0.0554\) & \(0.7273\) & \(0.8207 \pm 0.0249\) \\
      PR-AUC   & \(0.6596 \pm 0.0750\) & \(0.7704\) & \(0.9066 \pm 0.0200\) \\
      \hline
    \end{tabular}
  \end{minipage}
  \par
  \vspace{1em}
  \par

  \begin{minipage}{\textwidth}
    \begin{tabular}{c|c|c|c}
      \hline
      \textbf{Metric} & \textbf{Random Forest} & \textbf{SVM} & \textbf{Logistic Regression} \\
      \hline
      Accuracy & \(0.8292 \pm 0.0235\) & \(0.8292 \pm 0.0237\) & \(0.8044 \pm 0.0256\) \\
      F1       & \(0.8291 \pm 0.0235\) & \(0.8291 \pm 0.0237\) & \(0.8044 \pm 0.0256\) \\
      PR-AUC   & \(0.9173 \pm 0.0181\) & \(0.9142 \pm 0.0183\) & \(0.8889 \pm 0.0216\) \\
\bottomrule[0.8pt]
    \end{tabular}
  \end{minipage}
\end{table}

\begin{table}[htbp]
\centering
  \caption{Performance comparison for Clearance\_Hepatocyte\_AZ task.}
  \label{table_hepatocyte}
  \begin{minipage}{\textwidth}
    \begin{tabular}{c|c|c|c}
      \toprule[0.8pt]
      \textbf{Metric} & \textbf{QML average} & \textbf{QML best} & \textbf{XGBoost} \\
      \hline
      MSE      & \(0.4552 \pm 0.0159\) & \(0.4190\) & \(0.5335 \pm 0.0491\) \\
      Spearman & \(--\) & \(-0.3307\) & \(0.2184 \pm 0.0589\) \\
      \hline
    \end{tabular}
  \end{minipage}
  \par
  \vspace{1em}
  \par

  \begin{minipage}{\textwidth}
    \begin{tabular}{c|c|c}
      \hline
      \textbf{Metric} & \textbf{Random Forest} & \textbf{SVM} \\
      \hline
      MSE      & \(0.4467 \pm 0.0405\) & \(0.4123 \pm 0.0461\) \\
      Spearman & \(0.2479 \pm 0.0597\) & \(0.3327 \pm 0.0544\) \\
\bottomrule[0.8pt]
    \end{tabular}
  \end{minipage}
\end{table}

\begin{table}[htbp]
\centering
  \caption{Performance comparison for Clearance\_Microsome\_AZ task.}
  \label{table_microsome}
  \begin{minipage}{\textwidth}
    \begin{tabular}{c|c|c|c}
      \toprule[0.8pt]
      \textbf{Metric} & \textbf{QML average} & \textbf{QML best} & \textbf{XGBoost} \\
      \hline
      MSE      & \(0.4202 \pm 0.0297\) & \(0.3470\) & \(0.2775 \pm 0.0401\) \\
      Spearman & \(--\) & \(-0.4406\) & \(0.4643 \pm 0.0565\) \\
      \hline
    \end{tabular}
  \end{minipage}
  \par
  \vspace{1em}
  \par

  \begin{minipage}{\textwidth}
    \begin{tabular}{c|c|c}
      \hline
      \textbf{Metric} & \textbf{Random Forest} & \textbf{SVM} \\
      \hline
      MSE      & \(0.2756 \pm 0.0387\) & \(0.2848 \pm 0.0433\) \\
      Spearman & \(0.4870 \pm 0.0524\) & \(0.4571 \pm 0.0535\) \\
\bottomrule[0.8pt]
    \end{tabular}
  \end{minipage}
\end{table}

\begin{table}[htbp]
\centering
  \caption{Performance comparison for Half\_Life\_Obach task.}
  \label{table_halflife}
  \begin{minipage}{\textwidth}
    \begin{tabular}{c|c|c|c}
      \toprule[0.8pt]
      \textbf{Metric} & \textbf{QML average} & \textbf{QML best} & \textbf{XGBoost} \\
      \hline
      MSE      & \(0.2788 \pm 0.0757\) & \(0.1262\) & \(0.0197 \pm 0.0132\) \\
      Spearman & \(--\) & \(-0.4703\) & \(0.1736 \pm 0.0945\) \\
      \hline
    \end{tabular}
  \end{minipage}
  \par
  \vspace{1em}
  \par

  \begin{minipage}{\textwidth}
    \begin{tabular}{c|c|c}
      \hline
      \textbf{Metric} & \textbf{Random Forest} & \textbf{SVM} \\
      \hline
      MSE      & \(0.0105 \pm 0.0052\) & \(0.0063 \pm 0.0006\) \\
      Spearman & \(0.3450 \pm 0.0866\) & \(0.1401 \pm 0.0870\) \\
\bottomrule[0.8pt]
    \end{tabular}
  \end{minipage}
\end{table}

\begin{table}[htbp]
\centering
  \caption{Performance comparison for VDss\_Lombardo task.}
  \label{table_vdss}
  \begin{minipage}{\textwidth}
    \begin{tabular}{c|c|c|c}
      \toprule[0.8pt]
      \textbf{Metric} & \textbf{QML average} & \textbf{QML best} & \textbf{XGBoost} \\
      \hline
      MSE      & \(0.3073 \pm 0.0521\) & \(0.1412\) & \(0.0141 \pm 0.0128\) \\
      Spearman & \(--\) & \(-0.4651\) & \(0.2883 \pm 0.0674\) \\
      \hline
    \end{tabular}
  \end{minipage}
  \par
  \vspace{1em}
  \par

  \begin{minipage}{\textwidth}
    \begin{tabular}{c|c|c}
      \hline
      \textbf{Metric} & \textbf{Random Forest} & \textbf{SVM} \\
      \hline
      MSE      & \(0.0030 \pm 0.0024\) & \(0.0033 \pm 0.0002\) \\
      Spearman & \(0.3992 \pm 0.0600\) & \(0.2056 \pm 0.0649\) \\
\bottomrule[0.8pt]
    \end{tabular}
  \end{minipage}
\end{table}

\begin{table}[htbp]
\centering
  \caption{Real Hardware vs Simulation comparison for CYP2C9\_Substrate\_CarbonMangels.}
  \label{table_cyp2c9_hw}
  \begin{minipage}{\textwidth}
    \begin{tabular}{c|c|c|c}
      \toprule[0.8pt]
      \textbf{Metric} & \textbf{Simulation Max} & \textbf{Simulation Mean} & \textbf{Real HW Max} \\
      \hline
      Accuracy & \(0.741\) & \(0.668 \pm 0.059\) & \(0.719\) \\
      F1       & \(0.469\) & \(0.263 \pm 0.137\) & \(0.455\) \\
      PR-AUC   & \(0.430\) & \(0.339 \pm 0.043\) & \(0.343\) \\
      MSE      & \(1.013\) & \(0.929 \pm 0.049\) & \(1.038\) \\
      \hline
    \end{tabular}
  \end{minipage}
  \par
  \vspace{1em}
  \par

  \begin{minipage}{\textwidth}
    \begin{tabular}{c|c}
      \hline
      \textbf{Metric} & \textbf{Real HW Mean} \\
      \hline
      Accuracy & \(0.541 \pm 0.098\) \\
      F1       & \(0.293 \pm 0.111\) \\
      PR-AUC   & \(0.278 \pm 0.035\) \\
      MSE      & \(0.988 \pm 0.037\) \\
\bottomrule[0.8pt]
    \end{tabular}
  \end{minipage}
\end{table}

\begin{table}[htbp]
\centering
  \caption{Real Hardware vs Simulation comparison for CYP2D6\_Substrate\_CarbonMangels.}
  \label{table_cyp2d6_hw}
  \begin{minipage}{\textwidth}
    \begin{tabular}{c|c|c|c}
      \toprule[0.8pt]
      \textbf{Metric} & \textbf{Simulation Max} & \textbf{Simulation Mean} & \textbf{Real HW Max} \\
      \hline
      Accuracy & \(0.733\) & \(0.687 \pm 0.027\) & \(0.704\) \\
      F1       & \(0.540\) & \(0.257 \pm 0.140\) & \(0.497\) \\
      PR-AUC   & \(0.517\) & \(0.419 \pm 0.049\) & \(0.444\) \\
      MSE      & \(0.972\) & \(0.887 \pm 0.041\) & \(1.159\) \\
      \hline
    \end{tabular}
  \end{minipage}
  \par
  \vspace{1em}
  \par

  \begin{minipage}{\textwidth}
    \begin{tabular}{c|c}
      \hline
      \textbf{Metric} & \textbf{Real HW Mean} \\
      \hline
      Accuracy & \(0.576 \pm 0.097\) \\
      F1       & \(0.314 \pm 0.134\) \\
      PR-AUC   & \(0.347 \pm 0.048\) \\
      MSE      & \(0.979 \pm 0.063\) \\
\bottomrule[0.8pt]
    \end{tabular}
  \end{minipage}
\end{table}

\begin{table}[htbp]
\centering
  \caption{Real Hardware vs Simulation comparison for BBB\_Martins.}
  \label{table_bbb_hw}
  \begin{minipage}{\textwidth}
    \begin{tabular}{c|c|c|c}
      \toprule[0.8pt]
      \textbf{Metric} & \textbf{Simulation Max} & \textbf{Simulation Mean} & \textbf{Real HW Max} \\
      \hline
      Accuracy & \(0.860\) & \(0.783 \pm 0.064\) & \(0.800\) \\
      F1       & \(0.920\) & \(0.868 \pm 0.050\) & \(0.889\) \\
      PR-AUC   & \(0.930\) & \(0.880 \pm 0.022\) & \(0.891\) \\
      MSE      & \(0.973\) & \(0.806 \pm 0.105\) & \(1.234\) \\
      \hline
    \end{tabular}
  \end{minipage}
  \par
  \vspace{1em}
  \par

  \begin{minipage}{\textwidth}
    \begin{tabular}{c|c}
      \hline
      \textbf{Metric} & \textbf{Real HW Mean} \\
      \hline
      Accuracy & \(0.494 \pm 0.205\) \\
      F1       & \(0.538 \pm 0.289\) \\
      PR-AUC   & \(0.831 \pm 0.027\) \\
      MSE      & \(0.992 \pm 0.132\) \\
\bottomrule[0.8pt]
    \end{tabular}
  \end{minipage}
\end{table}

\begin{table}[htbp]
\centering
  \caption{Real Hardware vs Simulation comparison for Bioavailability\_Ma.}
  \label{table_bioavail_hw}
  \begin{minipage}{\textwidth}
    \begin{tabular}{c|c|c|c}
      \toprule[0.8pt]
      \textbf{Metric} & \textbf{Simulation Max} & \textbf{Simulation Mean} & \textbf{Real HW Max} \\
      \hline
      Accuracy & \(0.635\) & \(0.540 \pm 0.034\) & \(0.758\) \\
      F1       & \(0.711\) & \(0.658 \pm 0.043\) & \(0.862\) \\
      PR-AUC   & \(0.668\) & \(0.523 \pm 0.062\) & \(0.851\) \\
      MSE      & \(1.126\) & \(1.042 \pm 0.046\) & \(1.029\) \\
      \hline
    \end{tabular}
  \end{minipage}
  \par
  \vspace{1em}
  \par

  \begin{minipage}{\textwidth}
    \begin{tabular}{c|c}
      \hline
      \textbf{Metric} & \textbf{Real HW Mean} \\
      \hline
      Accuracy & \(0.615 \pm 0.111\) \\
      F1       & \(0.724 \pm 0.131\) \\
      PR-AUC   & \(0.770 \pm 0.042\) \\
      MSE      & \(0.931 \pm 0.079\) \\
\bottomrule[0.8pt]
    \end{tabular}
  \end{minipage}
\end{table}

\begin{table}[htbp]
\centering
  \caption{Real Hardware vs Simulation comparison for Clearance\_Hepatocyte\_AZ.}
  \label{table_hepatocyte_hw}
  \begin{minipage}{\textwidth}
    \begin{tabular}{c|c|c|c}
      \toprule[0.8pt]
      \textbf{Metric} & \textbf{Simulation} & \textbf{Real Hardware} & \textbf{Difference (\%)} \\
      \hline
      Mean Loss & \(0.4552 \pm 0.0156\) & \(0.5685 \pm 0.0874\) & \(24.9\) \\
      Min Loss  & \(0.4190\) & \(0.4392\) & \(4.8\) \\
      Max Loss  & \(0.4745\) & \(0.7222\) & -- \\
      Median    & \(0.4519\) & \(0.5490\) & -- \\
\bottomrule[0.8pt]
    \end{tabular}
  \end{minipage}
\end{table}

\begin{table}[htbp]
\centering
  \caption{Real Hardware vs Simulation comparison for Clearance\_Microsome\_AZ dataset.}
  \label{table_microsome_hw}
  \begin{minipage}{\textwidth}
    \begin{tabular}{c|c|c|c}
      \toprule[0.8pt]
      \textbf{Metric} & \textbf{Simulation} & \textbf{Real Hardware} & \textbf{Difference (\%)} \\
      \hline
      Mean Loss & \(0.4202 \pm 0.0291\) & \(0.5779 \pm 0.1505\) & \(37.5\) \\
      Min Loss  & \(0.3470\) & \(0.3867\) & \(11.4\) \\
      Max Loss  & \(0.4570\) & \(0.8707\) & -- \\
      Median    & \(0.4265\) & \(0.4955\) & -- \\
\bottomrule[0.8pt]
    \end{tabular}
  \end{minipage}
\end{table}

\begin{table}[h]
\centering

\caption{Correlation summary across classification datasets. Spearman $R$ is computed between circuit score and test loss (MSE). We also report the best F1 achieved among evaluated circuits for each scoring method.}
\label{tab:appendix_corr_cls}
\small
\setlength{\tabcolsep}{6pt}
\begin{tabular}{lcccc}
\toprule

& \multicolumn{2}{c}{Baseline (Elivagar)} & \multicolumn{2}{c}{Ours (QCS-ADME)} \\
\cmidrule(lr){2-3} \cmidrule(lr){4-5}
Dataset & Spearman $R$ & F1 (best) & Spearman $R$ & F1 (best) \\
\midrule
\textbf{hERG}                             & 0.058  & 0.843 & \textbf{-0.276} & \textbf{0.852} \\
\textbf{BBB\_Martins}                     & 0.021  & 0.916 & \textbf{0.224}$^{*}$ & \textbf{0.920} \\
\textbf{Bioavailability\_Ma}              & -0.182 & 0.678 & \textbf{-0.427} & \textbf{0.711} \\
\textbf{HIA\_Hou}                         & 0.203  & 0.887 & \textbf{0.412} & \textbf{0.891} \\
\textbf{CYP2C9\_Substrate\_CarbonMangels} & 0.218     & 0.286    & 0.340             & 0.286              \\
\textbf{CYP2D6\_Substrate\_CarbonMangels} & 0.031     & 0.471    & -0.136             & \textbf{0.541}              \\
\textbf{DILI}                             & -0.339    & 0.756    & -0.159              & 0.705              \\
\textbf{Pgp\_Broccatelli}                 & -0.162     & 0.752    &  -0.052             & 0.719              \\
\bottomrule

\end{tabular}
\vspace{2pt}
\footnotesize{$^{*}$Statistically significant correlation ($p<0.05$).}

\end{table}

\end{appendices}
\end{document}